\documentclass[twocolumn]{aastex631}

\usepackage{graphicx}
\usepackage{soul}
\usepackage{tabularx}
\usepackage{amsmath}
\usepackage{array}
\usepackage{bm}
\usepackage{xcolor}
\usepackage{cleveref}
\usepackage{subfigure}
\usepackage{hyperref}
\usepackage[normalem]{ulem}
\usepackage{xcolor}

\begin{document}

\title{Baryon Acoustic Oscillations analyses with Density-Split Statistics}

\author[0000-0002-9855-2342]{Tengpeng Xu}
\affiliation{Key Laboratory for Computational Astrophysics, National Astronomical Observatories, Chinese Academy of Sciences\\
Beijing 100101, China}
\affiliation{College of Astronomy and Space Sciences, University of Chinese Academy of Sciences\\
Beijing, 100049, China}

\author[0000-0002-2128-866X]{Yan-Chuan Cai}
\affiliation{Institute for Astronomy, University of Edinburgh, Royal Observatory\\
Blackford Hill, Edinburgh, EH9 3HJ, UK \email{cai@roe.ac.uk}}

\author[0000-0001-8919-7409]{Yun Chen}
\affiliation{Key Laboratory for Computational Astrophysics, National Astronomical Observatories, Chinese Academy of Sciences\\
Beijing 100101, China \email{chenyun@bao.ac.cn}}
\affiliation{College of Astronomy and Space Sciences, University of Chinese Academy of Sciences\\
Beijing, 100049, China}

\author[0000-0002-2618-5790]{Mark Neyrinck}
\affiliation{Department of Physics and Astronomy, University of Denver, Denver, CO 80210, USA}
\affiliation{Blue Marble Space Institute of Science, Seattle, WA 98104, USA}

\author[0000-0002-9276-917X]{Liang Gao}
\affiliation{Key Laboratory for Computational Astrophysics, National Astronomical Observatories, Chinese Academy of Sciences\\
Beijing 100101, China}

\author[0000-0003-2153-7758]{Qiao Wang}
\affiliation{Key Laboratory for Computational Astrophysics, National Astronomical Observatories, Chinese Academy of Sciences\\
Beijing 100101, China}
\affiliation{College of Astronomy and Space Sciences, University of Chinese Academy of Sciences\\
Beijing, 100049, China}

\begin{abstract}

Accurate modeling for the evolution of the Baryon Acoustic Oscillations (BAO) is essential for using it as a standard ruler to probe cosmology. We explore the non-linearity of the BAO in different environments using the density-split statistics and compare them to the case of the conventional two-point correlation function (2PCF). We detect density-dependent shifts for the position of the BAO with respect to its linear version using halos from N-body simulations. Around low/high-densities, the scale of the BAO expands/contracts due to non-linear peculiar velocities. As the simulation evolves from redshift 1 to 0, the difference in the magnitude of the shifts between high- and low-dense regions increases from the sub-percent to the percent level. The width of the BAO around high density regions increases as the universe evolves, similar to the known broadening of the BAO in the 2PCF due to non-linear evolution. In contrast, the width is smaller and stable for low density regions. We discuss possible implications for the reconstructions of the BAO in light of our results.

\end{abstract}


\section{Introduction} 
\label{sec:intro}

Baryon Acoustic Oscillations (BAO) represents the sound waves in the primordial baryon-photon plasma before recombination due to the interactions between gravity, provided by the total matter, and the pressure, provided by baryons. The largest scale at which sound waves can propagate by the time of recombination defines the sound horizon. After recombination, the scale of the sound horizon is frozen into the baryon density fluctuations (and hence the matter density fluctuations) in comoving units, denoted as $r_{\rm d}\sim147$~Mpc$\sim 100$~Mpc/$h$. $r_{\rm d}$ represents the sound horizon at the drag epoch ($z_{\rm d}$), when baryons are released from the Compton drag of the photons, determined by the ratio of baryon to matter densities \citep{Eisenstein1998, HS1996}. The sound horizon has been measured precisely by WMAP and Planck through the temperature fluctuations of the cosmic microwave background \citep[e.g.][]{Bennett2013, Planck2018}. These acoustic oscillations, imprinted in the matter density with a comoving scale of $r_{\rm d}$, result in a peak in the two-point correlation function of matter, serving as one of the most precise standard rulers in cosmology for measuring cosmological distances. In the last few decades, generations of surveys such as 2dFGRS, 6dFGS, SDSS, BOSS, eBOSS and DESI had used it to set constraints on cosmological parameters \citep{Adame2025, Beutler2011, Eisenstein2005, Ross2015, Alam2017, Alam2021, Zarrouk2018}. 

However, the position of the BAO is not fixed after recombination in comoving coordinates, which we assume throughout. Non-linear growth of structure is expected to cause the location of the BAO to shift, typically at the sub-percent level; the width of the BAO is broadened relative to its initial state \citep[e.g.][]{EisensteinSeo2007, Smith2008, Crocce2008, Seo2008, Seo2010, Orban2011}. 
This causes a loss of information for the BAO measurement.
A common practice is to perform reconstructions by using the observed late-time positions of galaxies, estimate their displacements and move the galaxies back to their initial conditions \citep[e.g.][]{Eisenstein2007, Noh2009, Padmanabhan2009, Seo2010, White2015, Schmittfull2015, Wang2017, Yu2017, Hada2018, von_Hausegger2022, Nikakhtar2022, Nikakhtar2023, Chen2023, ChenXY2024, ChenXY2024arxiv, Chan2024}. Recently, \cite{Chen2024} showed that for DESI BAO measurements, the systematic error on the isotropic BAO scale can be reduced to the 0.1\% level with a proper reconstruction algorithm to remove non-linear effects. Indeed, reconstructions applied to idealised cubic-box simulations and redshift surveys with large continuous areas were shown to substantially reduce the broadening and nonlinear shifts of the BAO, but this may not always be the case in observations with complicated survey masks \citep[e.g.][]{Anderson2012}, or when the shot-noise level of the galaxy sample is high \citep{White2010,Wang2019, Seo2022}. 

In principle, if we can track the non-linear evolution of the BAO precisely, forward modeling can also be applied to extract cosmological information \citep[e.g.][]{Babic2022}. In addition, introducing environmental-dependence weights for the reconstruction by up-weighting under-dense environments has been demonstrated to help sharpening the peak of the BAO \citep{Achitouv2015}. It has also been shown that reconstructions sharpen the BAO around voids more than for the 2PCF of mock galaxies \citep{Zhao2020}. All these indicate potential benefits for understanding the environmental dependence for the non-linear evolution of the BAO, which is the focus of our study.

In light of the above, we take a slightly different approach to analyze the BAO. We will identify density fluctuations in the late-time Universe around which the scale of the BAO may have changed relative to its initial position, and analyze them separately. This was first proposed by \citet{Neyrinck2018} using the sliced correlation functions.
Specifically, we anticipate the scale of the BAO to contract around over-dense regions due to peculiar infalls, and expand around under-dense regions due to outflows \citep[e.g.][]{Neyrinck2018, S&Z2012}. These opposite trends of evolution drive the broadening the BAO in conventional two-point statistics, which received mixed contributions from both the over-dense and under-dense regions. Additionally, the contraction and expansion are not necessarily symmetric due to non-linearity, causing an overall shift for the scale of the BAO in 2PCF. The broadening and shift of the BAO are entangled in two-point statistics \citep[e.g.][]{PadmanabhanWhite2009}. By analysing the BAO around over-dense and under-dense regions separately, we may be able to disentangle the broadening and the shifts of the BAO.

In this paper, we will perform BAO analyses with halos from the {\sc Quijote} simulations \citep{Francisco2020}. We will use the density-split clustering (DS), or conditional correlation functions for the analysis~\citep{Paillas2021,Paillas2023}. 
We smooth the halo number density field with a top-hat window function, rank and split the smoothed densities. By doing this, we will identify the possible expansion and contraction of the BAO scale around different density environments, with the perspective of tracking the broadening of the BAO.
We will explore if BAO analyses with DS-clustering have any significant impact in the error budget compared to two-point statistics. We will address both the shifts and the broadening of the BAO together by performing error analyses. 

Previous studies have shown that the shift in the BAO scale is correlated with the linear bias of the tracer. Tracers with larger linear bias tend to exhibit a greater shift in the measured BAO scale \citep[e.g.][]{PadmanabhanWhite2009}. We will show that density-split clustering will introduce a linear bias covering a much wider range than halo bias does, with under-dense regions associated with a negative bias and over-dense regions with a positive bias. This allows us to explore the relationship between BAO shift and linear bias more generally.

During the preparation of our manuscript, a study showing a detection of the negative shift of the BAO (i.e., a contraction of the BAO scale) in Ly$\alpha$ forest from cosmological simulations was reported \citep{Sinigaglia2024}. The paper was submitted to the arXiv in the same month as ours, and was independent. We will show that with DS-clustering, we can detect both the contraction and expansion of the BAO scale. 

Our paper is organized as follows. In Section~\ref{sec:measurements} we introduce the {\sc Quijote} simulations, the halo catalogs used in this work, and describing the measurements of the density-split clustering. We present our models and their parameters for the correlation functions in Section~\ref{sec:models}. Our BAO analyses are presented in Section~\ref{sec:mcmc_fitting}. We provide discussions and the main conclusions in Section~\ref{sec:discussion}.

\section{DENSITY-SPLIT clustering measurements from SIMULATIONS}
\label{sec:measurements}
In this section, we introduce the set up of the density-split clustering analyses. This follows closely the setting presented in \citep{Paillas2023} for their full-shape analyses.
\subsection{The {\sc Quijote} simulation}
The {\sc Quijote} project \citep{Francisco2020} comprises a series of N-body simulations aimed at assessing the information content in cosmological observables, spanning a wide range of values around its fiducial cosmology, which is consistent with the latest Planck constraints. The parameters of the fiducial cosmology are set as follows: $\Omega_{\rm m}=0.3175$, $\Omega_{\rm b}=0.049$, $h=0.6711$, $n_s=0.9624$, $\sigma_8=0.834$, $M_{\nu}(\rm eV)=0$, and $w=-1$. There are 15,000 realizations of the fiducial cosmology encompassed in the {\sc Quijote} simulations, providing a substantial dataset for the computation of covariance matrices. Dark matter particles and halos catalogs obtained with a Friends-of-Friends algorithm \citep{Davis1985} are provided. In our work, we utilized the dark matter halo catalogs with a minimal halo mass of $1.31\times10^{13}~M_{\odot}/h$ from 1000 realizations of the simulations. We focus on the data at redshifts $z=0$, 0.5, and 1 where most galaxy redshift surveys can cover.

\subsection{Measurements of clustering}
\label{subsec:measurements}
\begin{figure}
    \includegraphics[width=\linewidth]{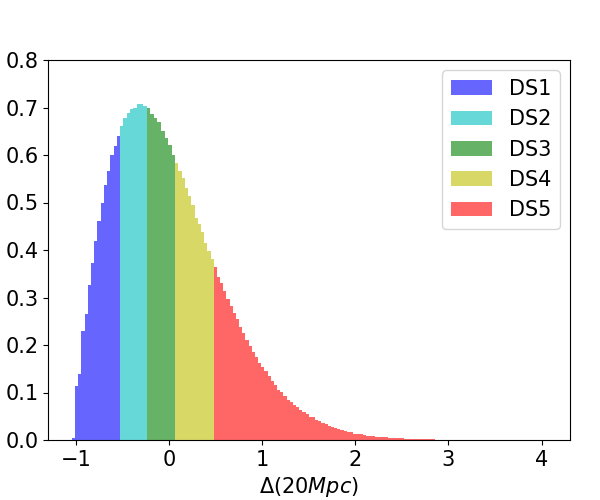}
    \caption{The normalized PDF of the halo number density contrast smoothed with a spherical top-hat window of the radius $R=20~$Mpc/$h$. Regions of increasing local densities, corresponding to ${\rm DS}_1$ to ${\rm DS}_5$ are shown in different colors. The auto-correlations of these DS centers, and their cross-correlations with halos are presented in Fig.~\ref{fig:haloxds_dsxds}. These are the basis for the density-split clustering analyses described in the paper. The measurement was conducted for one box of the {\sc Quijote} simulation at $z=0$.}
    \label{fig:pdf_ds}
\end{figure}

We follow the algorithm presented in \cite{Paillas2021, Paillas2023} to split the halo field into density bins. We apply spherical top-hat smoothing with a radius of $R = 20 h^{-1}\rm Mpc$ to the halo number density field, and measure the smoothed density at random locations (called DS centers, for centers of top-hat smoothing spheres) of the simulation box. Note that a DS center does not usually correspond to the location of a halo. The number of randoms in each box are set to be $M$ times the number of halos, $N_{\rm random}=M\times N_{\rm halos}$, where $M$ is the number of density bins we are going to use for the split. This is to ensure fair comparisons with the 2PCF \citep{Paillas2023}. This is equivalent to counting the number of halos around the random positions to measure the halo number density fluctuations within $R$: \begin{equation}
    \Delta(R) = \frac{\rho(<R)}{\bar{\rho}} - 1,
\end{equation}
where $\bar{\rho}$ is the mean halo density of the entire simulated box. The probability density function (PDF) of the density contrasts for a single realization at $z=0$ is shown in Fig.~\ref{fig:pdf_ds}. 

\begin{figure*}[ht!]
    \includegraphics[width=\linewidth]{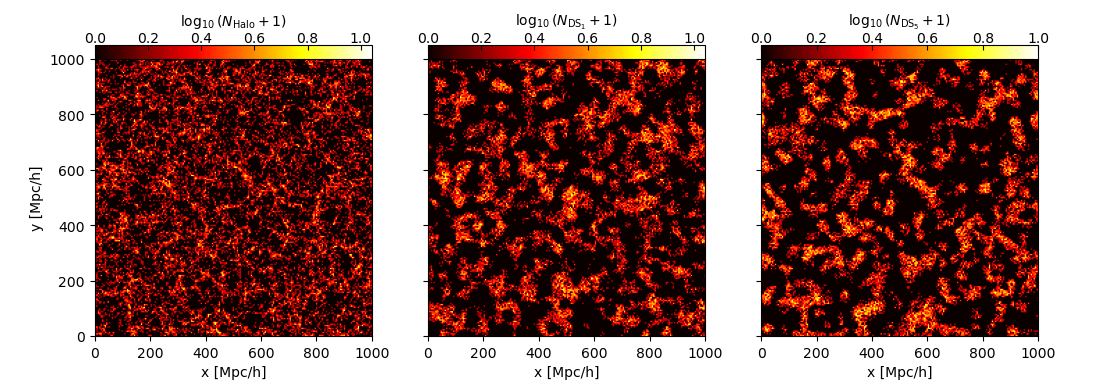}
    \caption{\textbf{Left}: A 2D slice of halo number counts from one realization of the {\sc Quijote} simulations. The thickness of the slice is 50~Mpc/$h$, with each pixel of the size of 5 Mpc/$h$ $\times$ 5 Mpc/$h$. \textbf{Middle}: Number counts of DS$_1$ centers (the most under-dense regions) selected from filtering the 3D halo number density field with a spherical top-hat filter of the radius of $20$Mpc/$h$ (see Section~\ref{subsec:measurements} and Fig.~\ref{fig:pdf_ds} for more details). \textbf{Right}: Similar to the middle panel but showing the number counts of DS$_5$ centers (the most over-dense regions). It can be observed that the distribution of DS$_1$ mostly corresponds to the void regions in the halo distribution, whereas the DS$_5$ distribution is largely complementary to that of DS$_1$, occupying high-dense regions.}
    \label{fig:ds_2d}
\end{figure*}

\begin{figure}[hb]
    \includegraphics[width=\linewidth]{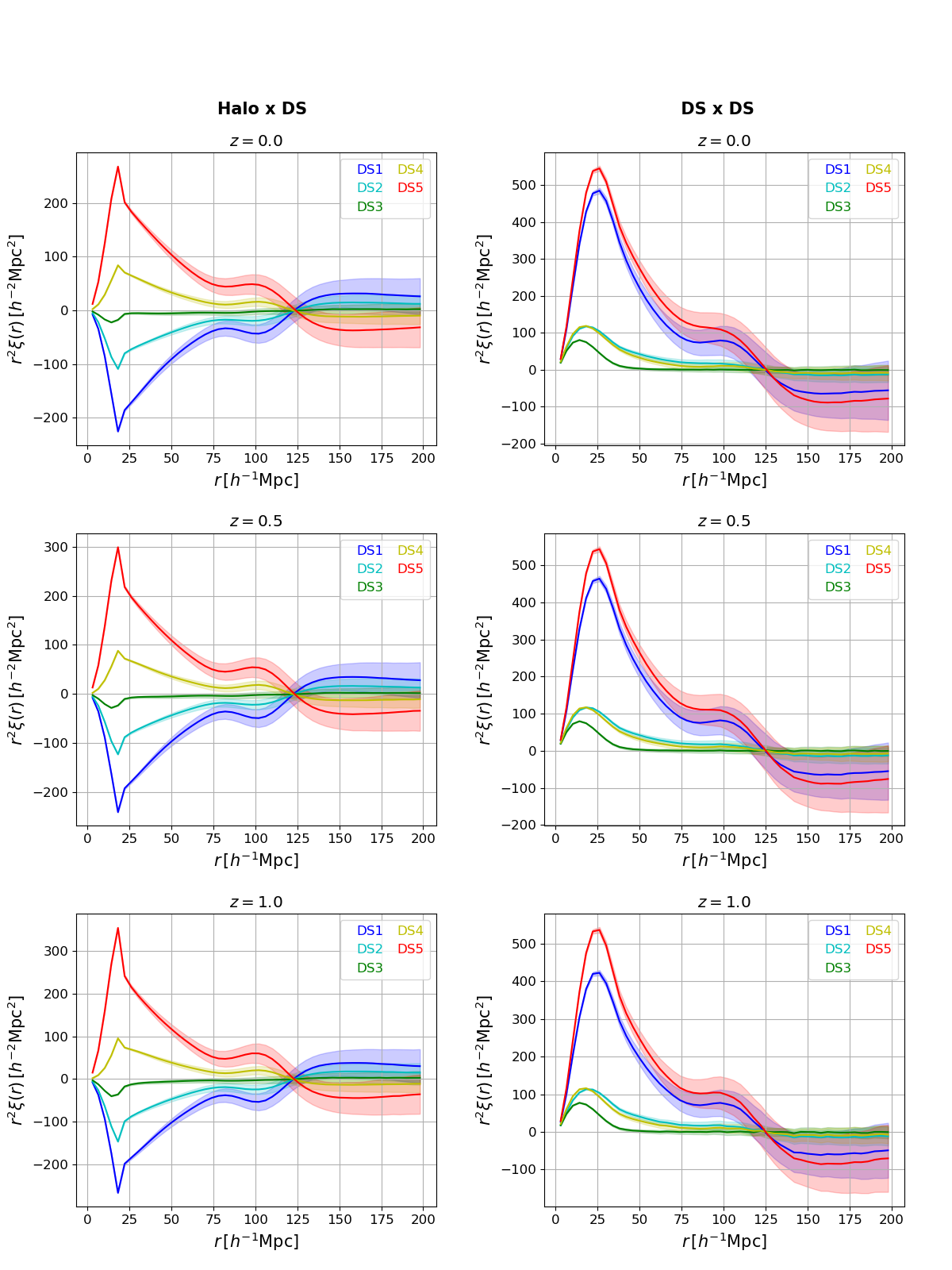}
    \caption{
    Measurements for the Halo-DS cross-correlation functions and the DS auto-correlation functions at redshifts $z=$ 0, 0.5 and 1. Solid lines are the averages from 1000 realizations of {\sc Quijote} simulations for the fiducial cosmology, with their errors shown in shaded regions.
    }
    \label{fig:haloxds_dsxds}
\end{figure}

Next, we divide the DS centers equally into $M$-bins and by default, five quintiles, naming them from $\rm DS_1$ to $\rm DS_5$ as indicated in Fig.~\ref{fig:pdf_ds}. Hence $\rm DS_1$ corresponds to under-dense regions, while $\rm DS_5$ corresponds to over-dense regions. We also repeat our analyses with $M=10$ with results presented in Appendix~\ref{apd:DS10}. Note that while the PDF is constructed by sampling the halo number density field at random locations, once split, the locations of $\rm DS_1$ and $\rm DS_5$ are no longer random. They are in fact strongly (anti-) biased, as seen in Fig.~\ref{fig:ds_2d} in terms of their spatial distribution, and in Fig.~\ref{fig:haloxds_dsxds} in terms of their auto-correlations and cross-correlations with halos.

Fig.~\ref{fig:ds_2d} display the distribution of halos, DS$_1$ and DS$_5$ centers in a two-dimensional slice with a thickness of $50$ Mpc$/h$ at redshift $z=0$ (corresponding to the three subplots from left to right, respectively). We can visually observe that the distribution of DS$_1$ mostly corresponds to the void regions in the halo distribution, whereas the DS$_5$ distribution is largely complementary to that of DS$_1$.

With these DS centers, we utilize the \texttt{pycorr} software package \citep{Sinha2019, Sinha2020} to calculate the cross-correlation functions between $\rm DS_1$-$\rm DS_5$ centers and dark matter halos for the 1000 realizations at $z=0, 0.5, 1.0$. We also compute the auto-correlation functions of $\rm DS_1$-$\rm DS_5$ centers. In the following text, we denote the cross-correlation functions as ``Halo $\times$ DS" or $\xi_i^{\rm qh}$, and the auto-correlation functions of ${\rm DS}_i$ as ``DS $\times$ DS" or $\xi_i^{\rm qq}$, where ``q" and ``h" are abbreviations for ``quintile" and ``halo", respectively. The measurements of these correlation functions are shown in Fig.~\ref{fig:haloxds_dsxds}.
We can see that $\xi_i^{\rm qh}$ measures the distribution of halos around the low-/high- density environments smoothed at 20~Mpc/$h$; and $\xi_i^{\rm qq}$ measures the auto-clustering of those low-/high- density regions. These are the basis of the DS-clustering which we will compare with the 2PCF in BAO analyses.

\section{THEORETICAL MODELS OF CORRELATION FUNCTIONS}
\label{sec:models}

For the model, we first utilize the \texttt{Nbodykit} package \citep{Nick2018} to generate the linear power spectrum and the corresponding correlation functions at $z=0, 0.5$ and 1.0 based on the same fiducial cosmology of the {\sc Quijote} simulations. We use the sideband algorithm~\citep{Kirkby2013} to decompose the correlation function into two parts: one that exclusively contains the BAO peak ($\xi_{\rm peak}(r)$) and another that represents the smooth portion without the BAO peak ($\xi_{\rm smooth}(r)$). We then Fourier transform these components to obtain the corresponding power spectra, $P_{\rm peak}(k)$, and $P_{\rm smooth}(k)$. Subsequently, we can calculate the quasi-linear power spectrum as:
\begin{equation}
    P_{\rm QL}(k; \sigma) = P_{\rm smooth}(k) + \exp(-\frac{k^2\sigma^2}{2})P_{\rm peak}(k),
\end{equation}
where $\sigma$ is the BAO peak's non-linear damping term (referred to as the BAO width parameter in the following text). Therefore, we can express the tracer biased power spectrum as given by
\begin{equation}
\label{eq:Pk}
    \hat{P}(k; b_1, b_2, \sigma) = b_1 b_2 P_{\rm QL}(k; \sigma),
\end{equation}
where $b_1$ and $b_2$ represent the biases of the two tracers of interest e.g., $b_{\rm halo}$ and $b_{\rm DSi}$ in the context of this study. We obtain the corresponding two-point correlation functions $\hat{\xi}(r; b_1, b_2, \sigma)$ through Fourier transformation of the model. Finally, 
we introduce the scale dilation parameter $\alpha$ for the distortions to the correlations due to assuming a cosmology which is different from the fiducial one \citep{AlcockPaczynski1979,Ballinger1996}. In general, the scale dilation parameters for the line-of-sight distance and the transverse distance are:
\begin{align}
\alpha_{\parallel}  &= \frac{[D_{\rm H}(z)/r_{\rm d}]}{[D_{\rm H}(z)/r_{\rm d}]_{\rm fid}}  \\
\alpha_{\perp}& = \frac{[D_{\rm M}(z)/r_{\rm d}]}{[D_{\rm M}(z)/r_{\rm d}]_{\rm fid}},
\end{align}
where $D_{\rm H}$ and $D_{\rm M}$ are the Hubble distance and the comoving angular diameter distance at $z$ respectively. These parameters is expected to provide the main cosmological constraints using the BAO.
In this work, we focus on the possible shifts of the BAO positions, and so we do not include redshift space distortions in our analyses. Therefore, the correlation functions are isotropic, so do the AP distortions, meaning that $\alpha_\parallel=\alpha_\perp$, where we denote as $\alpha$ in the following text. The theoretical model of the correlation function is given as:
\begin{equation}
    \xi_{\rm th}(r; b_1, b_2, \alpha, \sigma) = \hat{\xi}(\alpha*r; b_1, b_2, \sigma).
\end{equation}

Following the same notations introduced in Section~\ref{subsec:measurements}, ${\xi}_i^{\rm qh}(r; b_{{\rm halo|DS}_i}, b_{{\rm DS}_i}, \alpha^{\rm qh}_i, \sigma^{\rm qh}_i)$ represents the modeled cross-correlation function between dark matter halos and ${\rm DS}_i$ centers, where $b_{{\rm halo|DS}_i}, b_{{\rm DS}_i}, \alpha^{\rm qh}_i, \sigma^{\rm qh}_i$ are free parameters, and $i=1,2...5$. The same applies to the auto-correlation function of ${\rm DS}_i$ centers, ${\xi}_i^{\rm qq}(r; b_{{\rm DS}_i}, \alpha^{\rm qq}_i, \sigma^{\rm qq}_i)$, and the auto-correlation function of dark matter halos, $\xi^{\rm hh}(r; b_{\rm halo}, \alpha^{\rm hh}, \sigma^{\rm hh})$. It is noteworthy that, in this study, the halo bias in the cross-correlation functions between dark matter halos and ${\rm DS}_i$ centers (denoted as $b_{{\rm halo|DS}_i}$) and the halo bias in the auto-correlation functions of dark matter halos (denoted as $b_{\rm halo}$) are considered as distinct free parameters. This allows us to explore potential disparities between the halo bias around ${\rm DS}_i$ centers and the global halo bias.

The $\alpha$ parameter is a correction factor for the true cosmic distance relative to the fiducial cosmic distance. It also reflects deviations from the real position of the BAO relative to the reference cosmology. We treat $\alpha$ of different correlation functions to be independent free parameters, with $\alpha^{\rm hh}$ for the halo clustering, $\alpha^{\rm qh}_i$ for DS-Halo cross-correlations, and $\alpha^{\rm qq}_i$ for DS-DS auto-correlations. From the definition, when $\alpha>1$, the measured correlation function exhibits compression along the $r$ axis, and so the BAO scale is smaller than in the fiducial cosmology. Conversely, when $\alpha<1$, the opposite occurs. Only when $\alpha=1$, there is no deviation between the measured BAO peak position and that of the reference universe. 

In our analyses, the fiducial cosmology assumed is also the true cosmology, and so any deviation of $\alpha$ from 1 must come from non-linear evolution. We will call it the BAO shift parameter in the rest of the paper. The level of deviation for $\alpha$ for the 2PCF or power spectrum has been known to be at the sub-percent level at the late-time Universe \citep[e.g.][]{Crocce2008, Seo2008, Seo2010, Orban2011}. With density-split clustering, the BAO scale around different background densities may shift differently, and the possible deviation of $\alpha$ from 1 may become more obvious. In this work, we focus on the two most extreme densities -- ${\rm DS}_5$ and ${\rm DS}_1$, corresponding to the highest and lowest density fields. Guided by \citet{Neyrinck2018}, these largest density perturbations may cause the scale of the BAO to have the most significant shift. We will perform fitting with the model presented above with the measurements presented in section~\ref{sec:measurements} to investigate this. 

\section{MCMC fitting for BAO positions}
\label{sec:mcmc_fitting}

We employ the Markov Chain Monte Carlo (MCMC) method to constrain parameters in the correlation function model, utilizing the \texttt{emcee} software package \citep{Daniel2013}. For any of the correlation functions, $\xi^{\rm hh}$, $\xi^{\rm qh}_i$ or $\xi^{\rm qq}_i$, the log-likelihood can be computed by comparing the theoretical data vector $\bm{\xi_{\rm th}}$ with the measured data vector $\bm{\xi_{\rm sim}}$ obtained from simulations:
\begin{align}
    \ln \mathcal{L} &= -\frac{1}{2}\chi^2,\\
    \chi^2 &= (\bm{\xi_{\rm sim}}-\bm{\xi_{\rm th}})\mathbf{C}^{-1}(\bm{\xi_{\rm sim}}-\bm{\xi_{\rm th}})^\mathbf{T}.
\end{align}
For the data vector $\bm{\xi_{\rm sim}}$, we take the average over the measurements from the 1000 simulations. The covariance matrix $\mathbf{C}$ is also computed using the 1000 simulation boxes of the fiducial cosmology: 
\begin{equation}
    \mathbf{C} = \frac{1}{N_{\rm sim}-1}\sum^{N_{\rm sim}}_{k=1}(\bm{\xi_{{\rm sim}, k}} - \bm{\bar{\xi}_{\rm sim}})^\mathbf{T}(\bm{\xi_{{\rm sim}, k}} - \bm{\bar{\xi}_{\rm sim}}),
\end{equation}
where $N_{\rm sim}=1000$. Since the possible shift of the BAO peak position is expected to be small, we would like to reduce random errors as far as possible, but without introducing biases for the fitting. We scaled down the covariance matrix by a factor of 1/100, which corresponds to reducing the error bars of each data point to one-tenth of their original sizes. This is equivalent to increasing the volume of the simulated box by a factor of 100, namely, the new box has a side length of approximately $1~{\rm Gpc/}h\times 100^{1/3}\approx 4.6~{\rm Gpc/}h$. This volume may still be relevant for near future galaxy redshift surveys. For the fitting, we use the range of scales from 80 Mpc/$h$ to 180 Mpc/$h$ to encompasses the co-moving BAO scale  ($r\sim100~$Mpc/$h$), and make linear binning for the data with the bin-width of 4~Mpc/$h$. We have compiled the parameter constraint results for all the correlation functions described above in Table~\ref{tab:mcmc_results}. Next, we will proceed with a detailed analysis of each parameter.

\subsection{BAO constraints from halo-halo correlations}

\begin{figure}[htbp]
    \includegraphics[width=\linewidth]{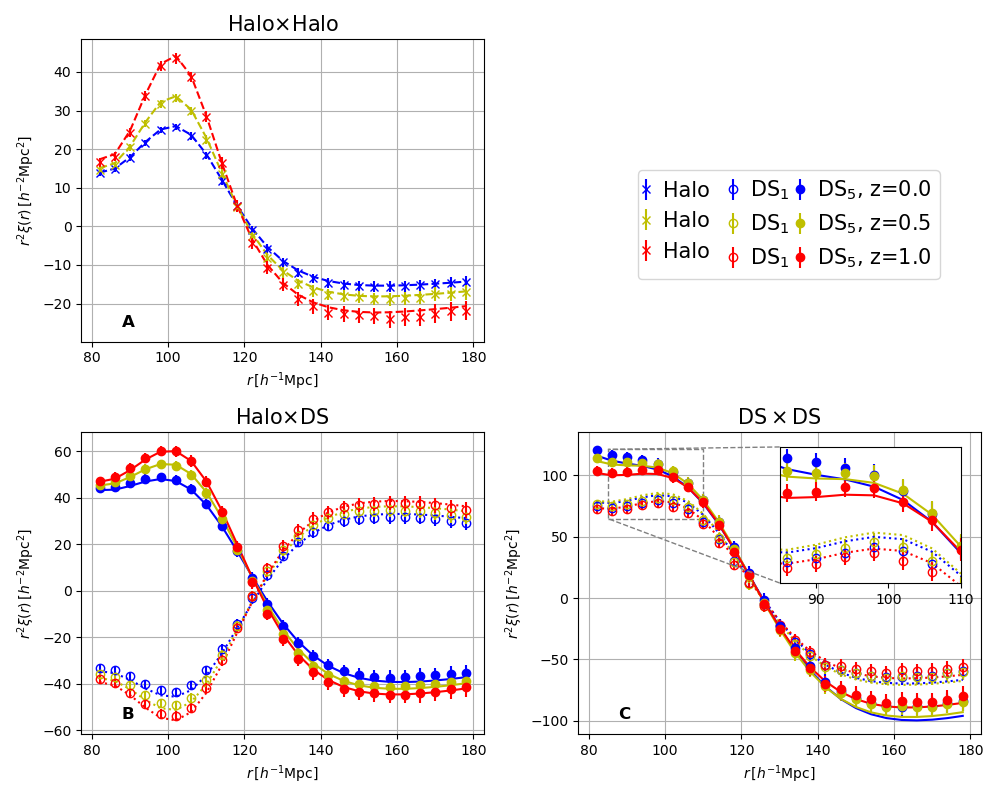}
    \caption{The best-fit results of the auto-correlation functions (Halo $\times$ Halo), as well as the joint-fit results of the cross-correlation functions (Halo $\times$ DS) and the auto-correlation functions (DS $\times$ DS) at $z=0.0$, 0.5 and 1.0. Points with error bars represent data obtained from simulated samples. The cross sign (`$\times$') denotes Halo $\times$ Halo correlation; solid and hollow circles are used to distinguish DS$_5$ and DS$_1$ for both Halo $\times$ DS and DS $\times$ DS correlation functions. Corresponding curves represent the best-fit models. Blue, yellow, and red lines correspond to $z=0.0$, 0.5 and 1.0, respectively.
    }
    \label{fig:halo_ds_curvefit}
\end{figure}

\begin{figure}
    \includegraphics[width=\linewidth]{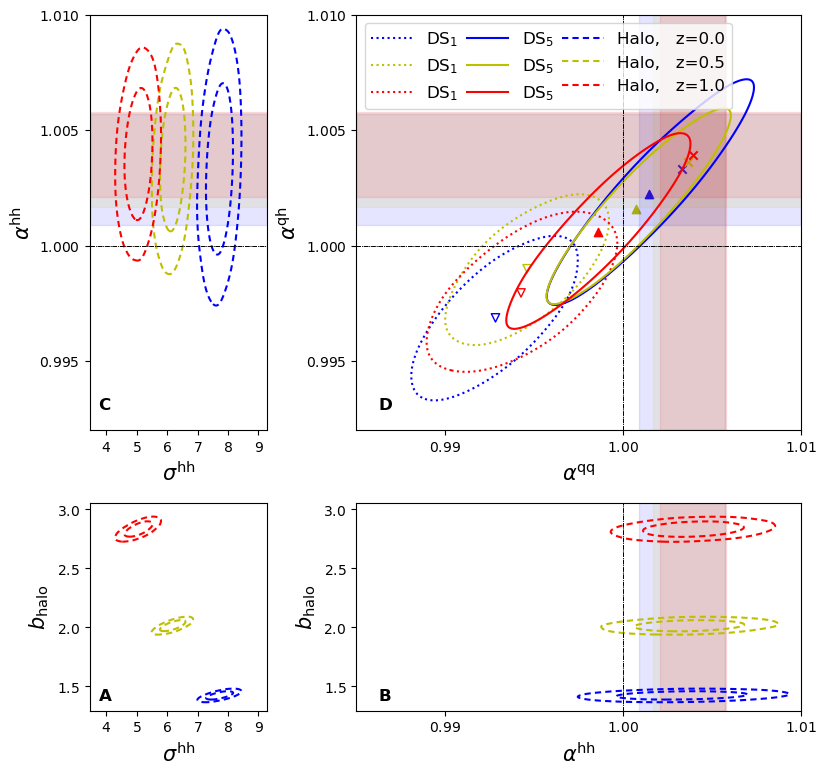}
    \caption{
    \textbf{Panels A, B, C} display the 2D posterior probability distributions of halo bias ($b_{\rm halo}$), the BAO shift parameter ($\alpha^{\rm hh}$), and the BAO width parameter ($\sigma^{\rm hh}$), obtained from MCMC fitting using halo auto-correlation functions. Large and small dashed contours represent the 68\% and 95\% confidence intervals, respectively.
    \textbf{Panel D} shows the 2D posterior probability distributions of the BAO shift parameters ($\alpha^{\rm qh}$ and $\alpha^{\rm qq}$) from joint constraints using DS $\times$ DS and Halo $\times$ DS correlation functions. Only the 68\% confidence contours are shown, with ${\rm DS}_1$ and ${\rm DS}_5$ distinguished by dotted and solid lines, respectively. Best-fit $\alpha^{\rm qh}$-$\alpha^{\rm qq}$ values for ${\rm DS}_1$ and ${\rm DS}_5$ are marked with hollow-downward triangles and solid-upward triangles,  respectively, while the best-fit $\alpha^{\rm hh}$ values are indicated with cross signs (``$\times$'').
    \textbf{Overall}: Blue, yellow, and red lines correspond to $z=0.0$, 0.5 and 1.0, respectively. The marginalized 68\% confidence intervals of the $\alpha^{\rm hh}$ parameter are shown as shaded bands in Panels B, C, and D for comparison. Additionally, both horizontal and vertical dashed black lines represent the unshifted alpha parameter ($\alpha = 1$).}
    \label{fig:halo_ds_alpha_mcmc}
\end{figure}

The fitted results for the halo-halo correlation functions (2PCF) at $z=0, 0.5$ and 1.0 are shown in Fig.~\ref{fig:halo_ds_curvefit}~(A). We can see that the model agrees very well with the measurements from simulations across the range of scales shown. Fig.~\ref{fig:halo_ds_alpha_mcmc}~(A, B, C) presents the constrained for $b_{\rm halo}, \alpha^{\rm hh}$ and $\sigma^{\rm hh}$. We can see that at these three redshifts, the central values of the alpha parameter are approximately 0.36\% greater than 1, and their differences from 1 all exceed the $1\sigma$ level. The fact that $\alpha^{\rm hh}>1$ suggests a contraction of the BAO scale relative to its linear version, consistent with previous studies \citep[e.g.][]{Crocce2008, Seo2008, Seo2010, Orban2011}. This is expected as halos with the bias greater than 1 occupied over-dense regions. Overall, the co-moving BAO scales are consistent with no significant evolution within the redshift range of 1 to 0. 

We can also see that the width of the BAO peak, as indicated by the parameter $\sigma^{\rm hh}$, changes from about 5.0 to 7.7 Mpc/$h$ from $z=1$ to $z=0$, a 50\% increase. This is expected from non-linear evolution \citep[e.g.][]{Seo2008}. 
In addition, the halo bias increases from $\sim 1.4$ to $\sim 2.8$ as the redshift increases from 0 to 1. This is expected as the density peaks corresponding to halos of the same low-mass cutoff become relatively rare at high-$z$. Additionally, the bias parameter $b_{\rm halo}$ appears uncorrelated with the $\alpha^{\rm hh}$, suggesting that the constraints on the BAO scale is unaffected by the amplitude of the clustering, as expected. All these indicate that our fitting pipeline works well for the 2PCF, and that there is no obvious shifts for the BAO scale in the 2PCF in this redshift range.

\begin{figure}
    \includegraphics[width=\linewidth]{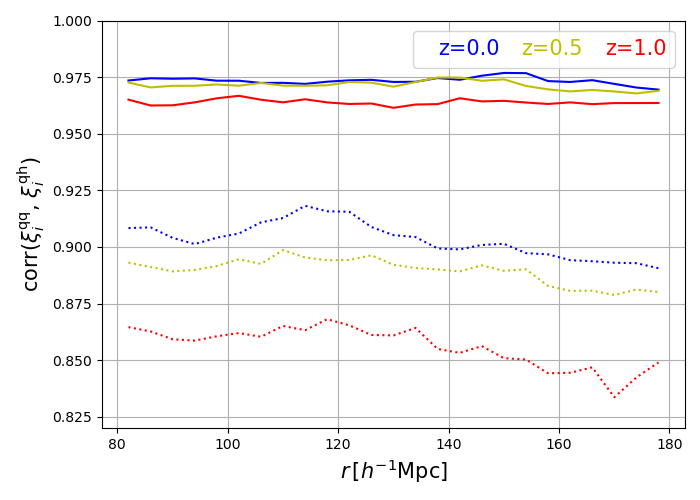}
    \caption{Comparing the correlation coefficients between the DS-DS and Halo-DS clustering at $z=0.0$, 0.5 and 1.0, where the dotted lines represent DS$_1$ (low densities) and the solid lines represent DS$_5$ (high densities). These are derived from the full covariance matrix of {[$\bf  \xi^{qq}, \xi^{qh}$}]. The coefficients for DS$_5$ are close to unity, suggesting that there is little gain of information when combining DS-DS with Halo-DS. The coefficients for DS$_1$ are typically smaller than those for DS$_5$, suggesting more room for improvement when combining DS-DS with Halo-DS for under-dense regions. }
    \label{fig:corr_qh_qq}
\end{figure}

\begin{figure}
    \includegraphics[width=\linewidth]{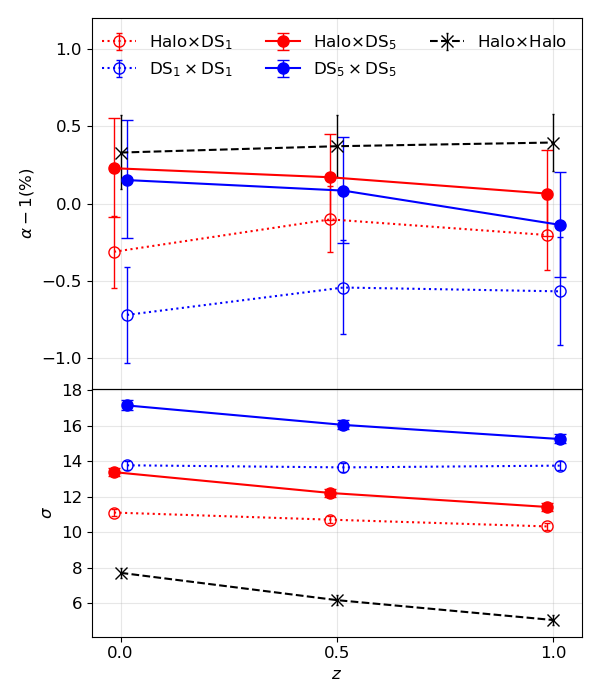}
    \caption{
    The marginalized constraints for the BAO shift parameter $\alpha$ and width parameter $\sigma$ at redshifts 0, 0.5, and 1. Black `$\times$' marks indicate the central values obtained from the halo auto-correlation function, while blue and red circles represent the central values obtained from the joint analysis of DS auto-correlation function and the Halo-DS cross-correlation function, respectively (hollow and solid circles distinguish DS$_1$ and DS$_5$). The 68\% confidence intervals for each parameter are displayed as error bars in the plot.}
    \label{fig:alpha_sigma_z}
\end{figure}

\begin{table}
    \centering
    \footnotesize
    \renewcommand{\arraystretch}{1.5} 
    \begin{tabular}{cccc}
    \hline \hline
        $z$ & $0.0$ & $0.5$ & $1.0$ \\ 
    \hline \hline
        $b_{\rm halo}$ & $1.425\pm{0.023}$ & $2.015\pm{0.031}$ & $2.832\pm{0.042}$ \\
        \hline
        $\alpha^{\rm hh}$ & $1.0033\pm{0.0025}$ & $1.0037\pm{0.0020}$ & $1.0039\pm{0.0019}$ \\
        \hline
        $\sigma^{\rm hh}$ & $7.72\pm{0.29}$ & $6.18\pm{0.27}$ & $5.06\pm{0.30}$ \\   
    \hline \hline
        $b_{\rm halo|DS_1}$ & $1.435\pm{0.023}$ & $2.041\pm{0.033}$ & $2.851\pm{0.048}$ \\
        \hline
        $b_{{\rm DS}_1}$ & $-3.02\pm{0.05}$ & $-3.96\pm{0.07}$ & $-4.83\pm{0.09}$ \\
        \hline
        $\alpha^{\rm qq}_1$ & $0.9928\pm{0.0031}$ & $0.9946\pm{0.0031}$ & $0.9943\pm{0.0035}$ \\
        \hline
        $\alpha^{\rm qh}_1$ & $0.9969\pm{0.0023}$ & $0.9990\pm{0.0021}$ & $0.9980\pm{0.0023}$ \\
        \hline
        $\sigma^{\rm qq}_1$ & $13.78\pm{0.25}$ & $13.66\pm{0.24}$ & $13.76\pm{0.28}$ \\
        \hline
        $\sigma^{\rm qh}_1$ & $11.12\pm{0.21}$ & $10.72\pm{0.20}$ & $10.34\pm{0.22}$ \\
    \hline \hline
        $b_{\rm halo|DS_5}$ & $1.423\pm{0.020}$ & $2.016\pm{0.028}$ & $2.817\pm{0.043}$ \\
        \hline
        $b_{{\rm DS}_5}$ & $3.66\pm{0.05}$ & $4.69\pm{0.07}$ & $5.68\pm{0.09}$ \\
        \hline
        $\alpha^{\rm qq}_5$ & $1.0015\pm{0.0039}$ & $1.0008\pm{0.0034}$ & $0.9986\pm{0.0034}$ \\
        \hline
        $\alpha^{\rm qh}_5$ & $1.0023\pm{0.0032}$ & $1.0017\pm{0.0028}$ & $1.0006\pm{0.0028}$ \\
        \hline
        $\sigma^{\rm qq}_5$ & $17.16\pm{0.27}$ & $16.06\pm{0.25}$ & $15.26\pm{0.26}$ \\
        \hline
        $\sigma^{\rm qh}_5$ & $13.39\pm{0.25}$ & $12.22\pm{0.22}$ & $11.44\pm{0.24}$ \\
    \hline \hline
    \end{tabular}
    \caption{The central values of parameters along with their $1\sigma$ errors from our BAO analyses fitted with $\xi^{\rm hh}$, $\xi^{\rm qq}$ and $\xi^{\rm qh}$, after marginalizing over other parameters. Note that $\xi^{\rm qq}$ and $\xi^{\rm qh}$ of the same DS bins are combined into a single data vector for the fitting. The errorbars present the statistical errors of a survey volume of 100~(Gpc/$h$)$^3$.}
    \label{tab:mcmc_results}
\end{table}

\subsection{BAO constraints from DS-clustering}
For density-split clustering, we combine ${\xi}_i^{\rm qq}$ and ${\xi}_i^{\rm qh}$ together to obtain joint constraints for the shift parameter. To account for the correlations between ${\xi}_i^{\rm qq}$ and ${\xi}_i^{\rm qh}$ at each redshift, we compute their joint covariance matrix and present the normalized correlation coefficients between ${\xi}_i^{\rm qq}$ and ${\xi}_i^{\rm qh}$ for both the DS$_1$ and DS$_{5}$ cases in Fig.~\ref{fig:corr_qh_qq}.
It can be seen from the figure that there is a strong correlation between ${\xi}_i^{\rm qq}$ and ${\xi}_i^{\rm qh}$. However, the correlation coefficients are significantly smaller for DS$_1$ than for DS$_{5}$. The cross-correlation functions  are relatively more independent from their auto-correlation functions around under-dense regions than the case around high-dense regions. This suggests that it is more valuable to combine ${\xi}_i^{\rm qq}$ and ${\xi}_i^{\rm qh}$ for DS$_1$ -- under-dense regions, than for DS$_5$ -- over-dense regions. 

We can see from Fig.~\ref{fig:halo_ds_curvefit}~(B, C) that the model provides excellent fits to the simulation measurements for both the auto-correlation functions (DS $\times$ DS) and the cross-correlation functions (Halo $\times$ DS) for ${\rm DS}_1$ and ${\rm DS}_5$ centers. 

\subsubsection{Biases}
 Both the theoretical model of the auto-correlation functions ${\xi}_i^{\rm qq}(r; b_{{\rm DS}_i}, \alpha^{\rm qq}_i, \sigma^{\rm qq}_i)$ and the cross-correlation functions ${\xi}_i^{\rm qh}(r; b_{{\rm halo|DS}_i}, b_{{\rm DS}_i}, \alpha^{\rm qh}_i, \sigma^{\rm qh}_i)$ have the parameter $b_{{\rm DS}_i}$. Their combination is able to break the degeneracy between $b_{\rm halo}$ and $b_{{\rm DS}_i}$, allowing independent measurements for the halo bias in different environmental densities. As shown in Table~\ref{tab:mcmc_results}, the biases of ${\rm DS}_1$ differ significantly from those of ${\rm DS}_5$, with $b_{{\rm DS}_1} \sim -3.0$ and $b_{{\rm DS}_5} \sim 3.6$ at $z=0$. These are significantly larger than the biases of the halos, suggesting that large-scale density fluctuations of these kinds are indeed rarer than the halos in our sample, but they are also more likely to be responsible for inducing non-linear evolution for the BAO. Furthermore, the biases for both ${\rm DS}_1$ and ${\rm DS}_5$ increase with redshift, similar to the evolution of the halo bias. Additionally, the halo biases measured in Halo-DS and DS-DS clustering are consistent with each other at the $1\sigma$ level between $\rm DS_1$ and $\rm DS_5$. Because the Halo-DS and DS-DS clustering measure the halo properties around DS centers, we can conclude that the biases of halos are consistent with each other around the low- and high-density environments.
They are also consistent with the global halo bias coming from the fitting with the 2PCF's. These results suggest that to the linear order, the halos of the same mass are statistically similar in different environments, once the environmental factor is accounted for by the DS biases. This agrees with the theoretical expectations \citep{Bardeen1986,Bond1991,Martino2009,Sheth1999} and findings from simulations \citep[e.g.][]{Alonso2015, Goh2019}.

\subsubsection{BAO scales}
Fig.~\ref{fig:halo_ds_alpha_mcmc}(D) presents the 2D posterior distributions
of the shift parameters $\alpha^{\rm qq}_i$ and $\alpha^{\rm qh}_i$ for ${\rm DS}_1$ and ${\rm DS}_5$, along with $\alpha^{\rm hh}$ at redshifts $z=0.0$, 0.5, and 1.0. Additionally, to facilitate more intuitive comparisons at different redshifts, we present the central values of the BAO shift parameters $\alpha$ with 1$\sigma$ error bars in Fig.~\ref{fig:alpha_sigma_z}.
Based on the information provided in Fig.~\ref{fig:halo_ds_alpha_mcmc}, Fig.~\ref{fig:alpha_sigma_z}, and Table~\ref{tab:mcmc_results}, the following insights can be intuitively gleaned: 

$\bullet$ Both the $\alpha^{\rm qq}$ and the $\alpha^{\rm qh}$ parameters for ${\rm DS}_5$ tend to be larger than those for ${\rm DS}_1$. This suggests that there are indeed contractions of the BAO scale around high-dense regions compared to low-dense regions.

$\bullet$ As the simulations evolves from redshift 1 to 0, the values of the $\alpha^{\rm qq}$ and $\alpha^{\rm qh}$ parameters for ${\rm DS}_5$ increase, whereas ${\rm DS}_1$ exhibits a mild opposite trend. Although the best-fit $\alpha^{\rm qq}$-$\alpha^{\rm qh}$ values for ${\rm DS}_1$ at $z=0.5$ are slightly higher than those at $z=1$, but they are consistent within the $1\sigma$ confidence level, which is probably due to the insignificant evolution of the BAO shifts between redshift 1 and 0.5. The evolution becomes more obvious when we split the density field into 10 DS bins (see Fig.~\ref{fig:halo_ds_alpha_mcmc_m10}). In the one-dimensional posterior distributions of the parameters at redshift 0, the difference between $\alpha^{\rm qh}_1$ and $\alpha^{\rm qh}_5$ nearly reaches $1\sigma$, while the difference between $\alpha^{\rm qq}_1$ and $\alpha^{\rm qq}_5$ exceeds $1\sigma$. The differences in their best-fit values are approximately 0.5\% and 1\%, respectively.
The fact that the $\alpha$ parameters around the low- and high- density regions evolve in opposite directions confirms the expansion/contraction of them in these two different environments due to non-linear evolution.

$\bullet$ The constrained results of parameters associated with the width of the BAO, $\sigma^{\rm qh}_i$ and $\sigma^{\rm qq}_i$, are larger than those of the 2PCF version $\sigma^{\rm hh}$ (bottom panel of Fig.~\ref{fig:alpha_sigma_z}). This is mainly due to the convolution of the top-hat window function when splitting the densities. In addition, the width of the BAO for $\rm DS_1$ is generally smaller than that for $\rm DS_5$ (dotted versus solid lines at the bottom panel of Fig.~\ref{fig:alpha_sigma_z}), which is consistent with the results of~\citep{Achitouv2015}. This difference is
possibly due to the smaller dynamical range of density for $\rm DS_1$ than $\rm DS_5$ (Fig.~\ref{fig:pdf_ds}). They also
exhibit different evolution between redshift one to zero. Both $\sigma^{\rm qh}_5$ and $\sigma^{\rm qq}_5$ increases with decreasing redshift, similar to the case of halo-halo clustering. However, both $\sigma^{\rm qh}_1$ and $\sigma^{\rm qq}_1$ remains stable across this redshift range. 

The increase for the width of the BAO with redshift in the 2PCF is expected from non-linear evolution. As halos reside on the peaks of the density field, they occupy a wide range of densities in the density PDF. The BAO's around different local densities evolve at different pace, resulting in a broadening of its width on average. In DS-clustering, $\rm DS_5$ takes the positive tail of the non-Gaussian PDF of the density field (see Fig.~\ref{fig:pdf_ds}), which also have a relatively wide dynamical range in density. Thus it can be expected that the width of the BAO around $\rm DS_5$ centers follows a similar evolution trend as that of the 2PCF. For $\rm DS_1$, however, their local densities have a narrow dynamical range (see Fig.~\ref{fig:pdf_ds}). The density perturbations are likely to evolve at a similar pace i.e., BAO's around under-dense regions will expand by a similar rate as they evolve. 
Their widths will therefore remain stable around these under-dense regions.

In summary, for DS-clustering, BAO's around under- and over-dense regions evolve at different rates. By analyzing them together, such as the case of the 2PCF, the situations in low- and high-dense regions will be mixed up. The expansion from under-dense regions and the contraction from high-dense regions will act together to broaden the width of the BAO.

\subsection{Understanding the shifts of the BAO}

The shift of the BAO scale measured by 2PCF, characterized by $\Delta\alpha\equiv\alpha-1$, was shown to be related to the bias parameters of tracers \citep{PadmanabhanWhite2009, S&Z2012}, 
\begin{equation}
    \Delta\alpha\propto(1+\frac{70}{47}\frac{b_2}{b_1})D(z)^2,
    \label{eqt: bao_shift}
\end{equation}
with the linear and quadratic bias parameters being $b_1$ and $b_2$, and $D(z)$ being the linear growth parameter. For DS-clustering, as we have shown in equation~\ref{eq:Pk}, the linear bias is the combination of $b_{\rm halo}$ and $b_{\rm DS_i}$, but we have seen that $b_{\rm halo}$ remains the same for ${\rm DS}_1$ and ${\rm DS}_5$, the only variable that matters for the linear bias is the bias related to density split i.e., $b_{\rm DS_i}$. So we can treat $b_{\rm DS_i}$ as the linear bias in this context.

The above model appears to be able to explain the shift of the BAO scale in DS-clustering as well. For the 2PCF of halos, in the regime where $b_1$ is large (e.g. $b_1>1$), $b_2$ increases faster than $b_1$ \citep[e.g.][]{PadmanabhanWhite2009, Modi2017}, thus we expect $\Delta\alpha$ to be positive, and $\Delta\alpha$ to increase with $b_1$ as seen in  Fig. 12 \& 13 of \citep{PadmanabhanWhite2009}. This is indeed the case for ${\rm DS}_5$ and ${\rm DS}_{10}$. From Fig.~\ref{fig:alpha_sigma_z} and Fig.~\ref{fig:alpha_sigma_z_m10} as well as Table~\ref{tab:mcmc_results} and Table~\ref{tab:mcmc_results_m10} at $z=0$, we do find that $\Delta\alpha$'s are positive for DS-clustering. Also, as the linear DS bias $b_{{\rm DS}_i}$ increases from 3.6 to 4.8 from ${\rm DS}_5$ to ${\rm DS}_{10}$ (high densities become more extreme), the amount of shift increases from 0.1-0.2\% to about 0.3-0.4\%. This agrees with the trend expected from the model.
For ${\rm DS}_1$ however, $b_{{\rm DS}_1}$ is strongly negative. This is a regime where the relationship between $b_1$ and $b_2$ have not been tested. If we were to extrapolate the $b_1-b_2$ relationship, we do expect $\Delta\alpha$ to become negative, and thus our measurement is consistent with the model. Also, as the linear DS bias becomes more negative going from the 5-density-bin to the 10-density-bin cases (low densities becomes more extreme), there is a hint of $\Delta\alpha$ becoming more negative. This again is consistent with the expectation of the model. 

In summary, as the linear DS bias going from negative to positive from ${\rm DS}_1$ to ${\rm DS}_5$, we have seen strong evidence that $\Delta\alpha$ changes from negative to positive. Despite the qualitative agreement with the model, and our confidence on $b_1$ for DS-clustering, we have not investigated $b_2$ for these cases, which is interesting itself for a future work. Therefore, we should be cautious about interpreting the qualitative agreement between our measurements and the model.

\section{DENSITY-SPLIT AFTER RECONSTRUCTION}
\label{sec:recon}

So far, we have seen clear evidence for the shifts of the BAO in DS-clustering. To validate if the shifts are indeed due to peculiar motions of halos around low and high dense regions, we perform the same analyses after reconstruction. 

We employ the typical reconstruction algorithm based on the Zel’dovich approximation \citep{Eisenstein2007}. A linear mapping between the late-time density and displacement field is assumed, therefore allowing us to use the halo number density field and the linear bias assumption to estimate the displacement field relative to the initial positions of halos. We can then use the displacement field to move the positions of the halos back to their initial conditions. Numerous studies \citep[e.g.][]{Eisenstein2007, Seo2008, Mehta2011, S&Z2012, Chen2024} have demonstrated that reconstruction can effectively remove most of the non-linear effects, thus sharpening the peak and reducing the shift of the BAO in two-point statistics. 

\begin{figure}
    \centering
    \includegraphics[width=\linewidth]{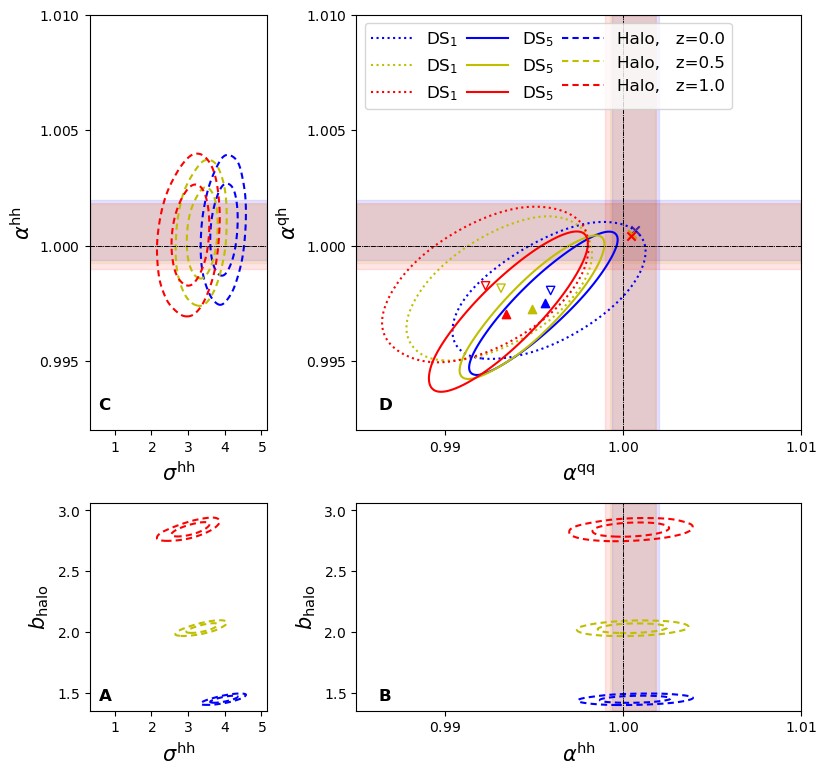}
    \caption{The constraint results of the Density Split and BAO analysis after reconstruction, follow the same details as shown in Fig.~\ref{fig:halo_ds_alpha_mcmc}.}
    \label{fig:mcmc_recon}
\end{figure}

\begin{figure}
    \centering
    \includegraphics[width=\linewidth]{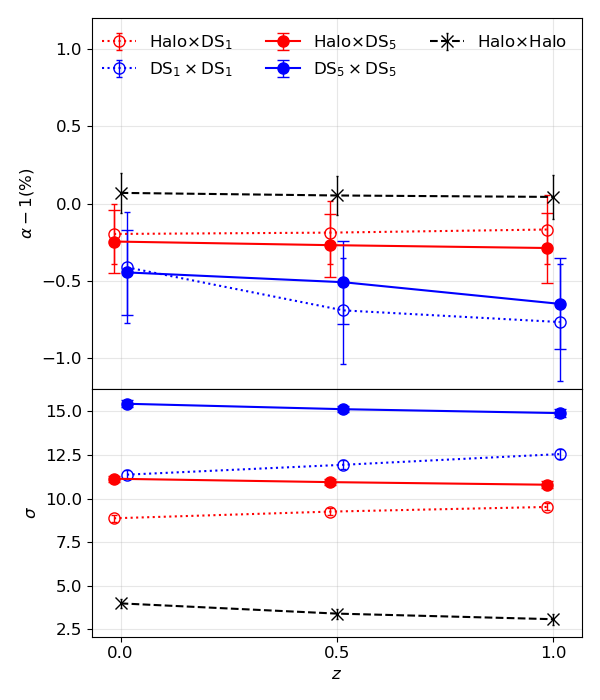}
    \caption{The constraint results of the Density Split and BAO analysis after reconstruction, follow the same details as shown in Fig.~\ref{fig:alpha_sigma_z}.}
    \label{fig:alpha_sigma_z_recon}
\end{figure}

We utilize the \texttt{pyrecon}\footnote{https://github.com/cosmodesi/pyrecon} package to reconstruct the positions of halos and then performed the same density-split BAO analyses on the reconstructed halo catalogs. The main results are shown in Fig.~\ref{fig:mcmc_recon} and Fig.~\ref{fig:alpha_sigma_z_recon}. 
Compared to the pre-reconstruction results shown in Fig.~\ref{fig:halo_ds_alpha_mcmc} and Fig.~\ref{fig:alpha_sigma_z}, we observe the following differences before and after the reconstruction: 

For the halo auto-correlation function, reconstruction effectively removes most of the BAO shift. However, although the BAO broadening (i.e., the $\sigma$ parameter) is significantly reduced, it is still not zero, indicating that the reconstruction method has not fully eliminated the nonlinear broadening of the BAO. This is expected with the Zel’dovich approximation.

For the DS auto-correlation functions and their cross-correlation with halos, the difference in BAO shift between low-dense regions (${\rm DS}_1$) and high-dense regions (${\rm DS}_5$) nearly disappears. Therefore, using the reconstructed peculiar velocities, the positions of the BAO have been largely re-installed in both high- and low-dense regions, though the residual nonlinear effects contribute to a remaining small shift of the BAO. In addition, the width of the BAO are reduced after reconstruction, and it shows little evolution. 

Therefore, we can conclude that peculiar velocities are indeed responsible for driving the shifts and broadening of the BAO in DS-clustering. With the spherical nature of the DS1 and DS5 regions on average, these results also indicate the possibility of modeling the shifts of the BAO with spherical dynamics, which we will explore in a future work.

\section{Discussion and conclusions}
\label{sec:discussion}

In this study, we have explored the possible deviation of the BAO scale in high-dense and low-dense regions due to non-linear evolution. We do so by analyzing the BAO with the density-split statistics using halos found in the {\sc Quijote} simulations. The density-split clustering starts from splitting the halo number density PDF smoothed at a 20 Mpc/$h$-scale into high- and low-dense regions, called DS centers. We then measure the auto-correlations of these DS centers and their cross-correlations with the halos. We also compare the results with the standard analysis of the BAO using 2PCF. The main conclusions drawn from our analyses are as the follows:

For the scales of the BAO, using the the auto-correlation functions of DS centers and the cross-correlation functions between halos and DS centers, we observed a contraction (stretching) of the BAO scale in high-dense (low-dense) regions relative to its linear version, while the scale of the BAO in the auto-correlation function of halos remains stable in the same redshift range. At redshift zero, the difference in the BAO scale between high- and low-dense regions is approximately 1\%.

Two-point statistics can be seen as the composition of the whole spectrum of cross-correlations in DS-clustering. The sub-percent level shifts for the BAO in 2PCF is the sum of all the effects around high- and low-dense regions in DS-clustering. Due to the opposite direction for the shifts of the BAO scale around high- and low-densities, a non-linear transformation for the density field which up-weight low-densities may compensate the contraction effect of the BAO around high-densities, and therefore reduce the shifts of the BAO in two-point statistics \citep{McCullagh2013}; a similar finding using the power spectrum, for general features, was explored by \citet{NeyrinckYang2013}.

For the width of the BAO, with density-split clustering, we find it to be larger than its 2PCF counterpart due to a convolution of a smoothing window function. The width is broadened by $\sim50\%$ between redshift 1 to 0 for the 2PCF. Similar evolution is seen for DS-clustering for the high-dense regions. For the highest density bin in DS-clustering, DS$_5$, it encloses the positive tail of the density PDF, which is more extended than the lowest density bin. The wide dynamical range of the local density, and the relatively strong non-linearity around these high-densities may explain the relatively more rapid evolution for the width of the BAO around them. The width of the BAO does not appear to evolve for low-dense regions, possibly due to having a similar evolution trend for the BAO around a narrow range of low-dense environments i.e., the evolution of BAO's around voids may be relatively synchronized. This may also explains the fact that reconstructions by up-weighting low-dense environments is beneficial in sharpening the BAO feature \citep{Achitouv2015}, which is also seen in the reconstruction with mock galaxy catalogs \citep{Zhao2020}.  

Our results agree qualitatively with the findings of \citep{Neyrinck2018}, but the magnitudes of the shifts are smaller. This is mainly because we are looking at a more realistic case where the density field is split into 5 quintiles. The local densities for $\rm DS_1$ and $\rm DS_5$ are not very extreme. Their density contrasts $\delta$ differ from the mean by an order of 1. More significant shifts for the BAO scale is expected if one samples the density contract of $\delta\sim 10$ \citep{Neyrinck2018}. However, at the smoothing scale of 20 Mpc/$h$, density fluctuations of this amplitude are rare, and the measurements for the DS-clustering become noisy. With a smaller smoothing scale, one can increase the amplitude of density fluctuations while keeping the noise level down. However, having smaller scale fluctuations means weaker gravitational influence on the BAO, and hence smaller impacts on the BAO scale.  

Nevertheless, with the default smoothing scale of $20$Mpc/$h$, we have repeated the analyses with 10 DS bins. Results are shown in Appendix~\ref{apd:DS10}. In this case, $\rm DS_1$ and $\rm DS_{10}$ are more extreme low and high densities. We have indeed observe a more obvious trend for the expansion of the BAO scale for $\rm DS_1$, but the evolution for the BAO scale for $\rm DS_{10}$ seems diminishing. The difference between these low and high densities becomes more extreme, reaching $\sim 2$\% at $z=0$. Overall, the results from using 10 DS-bins remains qualitatively similar to the case of 5 bins. 

The contraction and expansion of the BAO scales around high- and low-dense environments agrees qualitatively with models predicting the shifts of the BAO in two-point statistics, \citep{PadmanabhanWhite2009, S&Z2012}, providing that the relationship between linear bias $b_1$ and quadratic bias $b_2$ is extrapolated to the regime where $b_1$ is negative. We have also found that reconstructions based on the Zel'dovich approximation significantly reduced the shifts and broadening of the BAO in DS-clustering. This confirms the physical scenario that the contraction and expansion of the BAO are indeed due to peculiar infalls and outflows around high- and low-dense regions.

In conclusion, we have detected clear deviations for the BAO features between the low- and high- density environments, both in terms of the shift of its scale and in its broadening. These results may provide theoretical grounds for better interpretations of the BAO beyond two-point statistics. Accurate modeling for the environmental dependents of these features, e.g. through density-split clustering, may allow us to better predict the non-linear behavior of the BAO, and thus extract more cosmological information. 

We caution that with a focus on understanding the physics, our analyses are conducted in real space, but we anticipate the main conclusions to be similar in redshift space. If we take the perspective that the scenario in redshift space is a future universe in one of the three dimensions, the evolution effects of the BAO we have seen may be even stronger in redshift space. Indeed, this was seen for the shifts of the BAO in terms of the 2PCF \citep{Seo2008}.

\begin{acknowledgments}
This work has been supported by the National Key Research and Development Program of China (Nos.\ 2022YFA1602903 and 2023YFB3002501), the National Natural Science Foundation of China (Nos.\ 11988101 and 12033008), the China Manned Space Project (No.\ CMS-CSST-2021-A01), and the K.\ C.\ Wong Education Foundation. YC acknowledges the support of the UK Royal Society through a University Research Fellowship. WQ acknowledges the 
supported by the Strategic Priority Research Program of Chinese Academy of Sciences, Grant No.XDB0500203.
For the purpose of open access, the author has applied a Creative Commons Attribution (CC BY) license to any Author Accepted Manuscript version arising from this submission. 
\end{acknowledgments}

\software{
\texttt{densitysplit}~\citep{Paillas2021},
\texttt{pycorr}~\citep{Sinha2019, Sinha2020},
\texttt{Nbodykit}~\citep{Nick2018},
\texttt{pyrecon}~(https://github.com/cosmodesi/pyrecon),
\texttt{emcee}~\citep{Daniel2013},
\texttt{getdist}~\citep{Lewis2019}
}

\appendix

\section{the density-split clustering with 10 bins}
\label{apd:DS10}

\begin{figure}[ht!]
    \centering
    \subfigure[For the 10 density bin case, comparing the correlation coefficients between the DS-DS and Halo-DS clustering at $z=0.0$, 0.5 and 1.0, where the dotted lines represent DS$_1$ and the solid lines represent DS$_{10}$. ]{
        \includegraphics[width=0.45\textwidth]{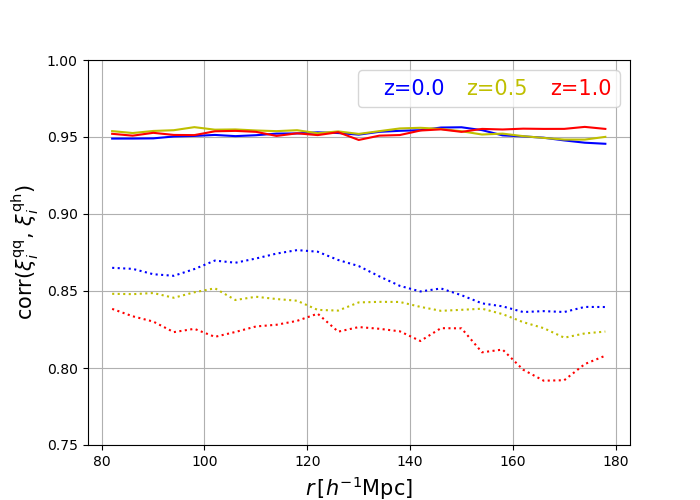}
        \label{fig:corr_qh_qq_m10}
    }
    \hfill
    \subfigure[For the case of 10 density bins, the best-fit results of the auto-correlation functions (Halo $\times$ Halo), as well as the joint fit results of the cross-correlation functions (Halo $\times$ DS) and the auto-correlation functions (DS $\times$ DS) at $z=0.0$, 0.5 and 1.0. ]{
        \includegraphics[width=0.45\textwidth]{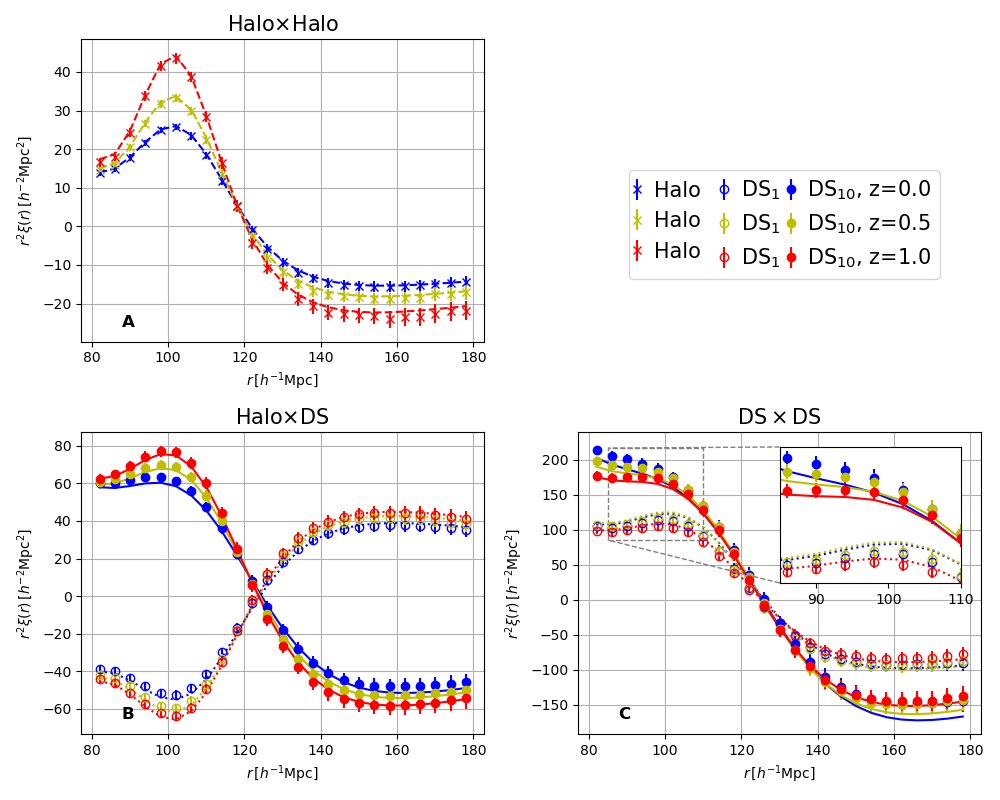}
        \label{fig:halo_ds_curvefit_m10}
    }
    \vfill
    \subfigure[For the 10 density bin case, the 2D probability distributions of the BAO shift parameters ($\alpha^{\rm qh}$ and $\alpha^{\rm qq}$) from joint constraints using DS $\times$ DS and Halo $\times$ DS correlation functions, with the details similar to Fig.~\ref{fig:halo_ds_alpha_mcmc}. ]{
        \includegraphics[width=0.45\textwidth]{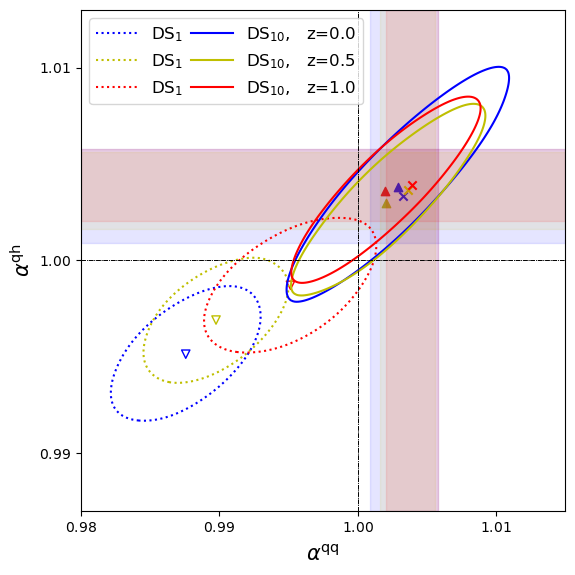}
        \label{fig:halo_ds_alpha_mcmc_m10}
    }
    \hfill
    \subfigure[For the 10 density bin case, the marginalized constraints for the BAO shift parameter $\alpha$ and width parameter $\sigma$ at redshifts 0, 0.5, and 1. ]{
        \includegraphics[width=0.45\textwidth, height=9.3cm]{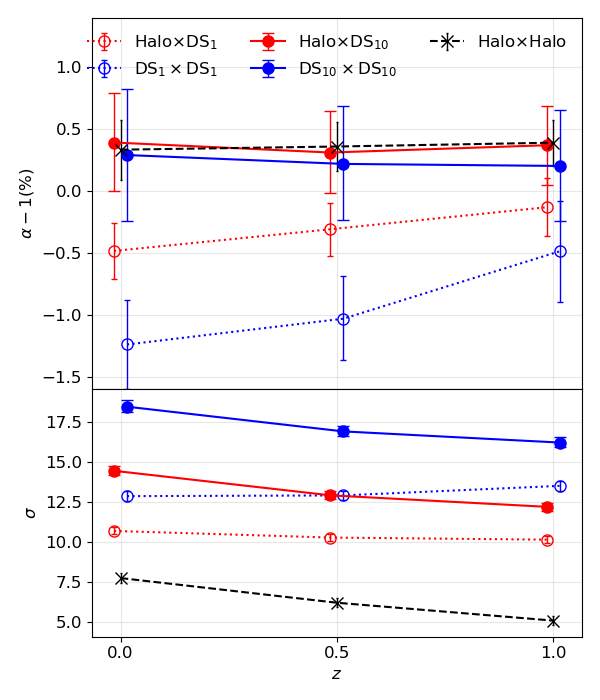}
        \label{fig:alpha_sigma_z_m10}
    }
    \caption{Main results of density-split analysis with 10 density bins.}
    \label{fig:both_figures}
\end{figure}

We repeat our analyses by splitting the halo density field into 10 density bins, with results shown in~\Cref{fig:corr_qh_qq_m10,fig:halo_ds_curvefit_m10,fig:halo_ds_alpha_mcmc_m10,fig:alpha_sigma_z_m10}. These can be compared with ~\Cref{fig:corr_qh_qq,fig:halo_ds_curvefit,fig:halo_ds_alpha_mcmc,fig:alpha_sigma_z} for the default case of having 5 density bins presented in the main part of the paper. The parameter constraint results are compiled in Table~\ref{tab:mcmc_results_m10}, which can be compared with the 5-density-bins case (Table~\ref{tab:mcmc_results}).

$\bullet$ Comparing Fig.~\ref{fig:corr_qh_qq} and Fig.~\ref{fig:corr_qh_qq_m10}, we can observe that in the case of 10 bins, the correlation coefficients between the halo-DS cross-correlation function and the DS auto-correlation function are all lower than those in the case of 5 bins. This indicates that as the density bins become finer, the information contained in each DS bin becomes more independent from the entire halo density field.

$\bullet$ Comparing Fig.~\ref{fig:halo_ds_curvefit} and Fig.~\ref{fig:halo_ds_curvefit_m10}, we can see that the correlation function model used in this work successfully fits the DS correlation functions measured from the simulation data for the 10-bin case as well as for the 5-bin case. 

$\bullet$ Comparing Fig.~\ref{fig:halo_ds_alpha_mcmc} with Fig.~\ref{fig:halo_ds_alpha_mcmc_m10} or the upper panels of Fig.~\ref{fig:alpha_sigma_z} and Fig.~\ref{fig:alpha_sigma_z_m10}, it becomes clear that in the 10-bin case, while the BAO shift in the highest dense region (${\rm DS}_{10}$) remains almost unchanged with redshift, the BAO shift in the lowest density region (${\rm DS}_1$) becomes more pronounced with redshift evolution. This results in a greater difference between high-dense and low-dense regions compared to the 5-bin case. This is expected as the more extreme the density fluctuations are, the stronger the infalls and outflows are stronger.

$\bullet$ Comparing the lower panels of Fig.~\ref{fig:alpha_sigma_z} and Fig.~\ref{fig:alpha_sigma_z_m10}, we can observe that the BAO width, represented by the $\sigma$ parameter, exhibits similar behavior in both the 5-bin and 10-bin cases. In high-density regions, the BAO width evolves with redshift in a similar manner as in the case of the halo 2PCF. In contrast, the BAO width in low-density regions shows little evolution with redshift.

In summary, as the density bins become more finely divided, the default results presented in the main paper for the shifts and broadening of the BAO with DS-clustering remains qualitatively similar, but the divergence between the low and high dense regions becomes more obvious. These indicates the robust of our analyses and the validity of the physical scenario presented in the main paper.

\begin{table}[hb!]
    \centering
    \footnotesize
    \renewcommand{\arraystretch}{1.15} 
    \begin{tabular}{cccc}
    \hline \hline
        $z$ & $0.0$ & $0.5$ & $1.0$ \\ 
    \hline \hline
        $b_{\rm halo}$ & $1.425\pm{0.023}$ & $2.015\pm{0.031}$ & $2.832\pm{0.043}$ \\
        \hline
        $\alpha^{\rm hh}$ & $1.0034\pm{0.0024}$ & $1.0036\pm{0.0020}$ & $1.0039\pm{0.0019}$ \\
        \hline
        $\sigma^{\rm hh}$ & $7.72\pm{0.29}$ & $6.18\pm{0.27}$ & $5.05\pm{0.30}$ \\    
    \hline \hline
        $b_{\rm halo|DS_1}$ & $1.426\pm{0.026}$ & $2.019\pm{0.036}$ & $2.839\pm{0.051}$ \\
        \hline
        $b_{{\rm DS}_1}$ & $-3.56\pm{0.07}$ & $-4.68\pm{0.09}$ & $-5.64\pm{0.11}$ \\
        \hline
        $\alpha^{\rm qq}_1$ & $0.9876\pm{0.0036}$ & $0.9897\pm{0.0035}$ & $0.9952\pm{0.0041}$ \\
        \hline
        $\alpha^{\rm qh}_1$ & $0.9952\pm{0.0023}$ & $0.9969\pm{0.0021}$ & $0.9987\pm{0.0023}$ \\
        \hline
        $\sigma^{\rm qq}_1$ & $12.85\pm{0.30}$ & $12.90\pm{0.30}$ & $13.50\pm{0.32}$ \\
        \hline
        $\sigma^{\rm qh}_1$ & $10.68\pm{0.22}$ & $10.26\pm{0.21}$ & $10.13\pm{0.22}$ \\
    \hline \hline
        $b_{\rm halo|DS_{10}}$ & $1.423\pm{0.021}$ & $2.001\pm{0.029}$ & $2.826\pm{0.042}$ \\
        \hline
        $b_{{\rm DS}_{10}}$ & $4.83\pm{0.07}$ & $6.10\pm{0.09}$ & $7.45\pm{0.12}$ \\
        \hline
        $\alpha^{\rm qq}_{10}$ & $1.0029\pm{0.0053}$ & $1.0022\pm{0.0046}$ & $1.0020\pm{0.0045}$ \\
        \hline
        $\alpha^{\rm qh}_{10}$ & $1.0039\pm{0.0040}$ & $1.0031\pm{0.0033}$ & $1.0037\pm{0.0032}$ \\
        \hline
        $\sigma^{\rm qq}_{10}$ & $18.46\pm{0.38}$ & $16.91\pm{0.32}$ & $16.22\pm{0.32}$ \\
        \hline
        $\sigma^{\rm qh}_{10}$ & $14.44\pm{+0.29}$ & $12.90\pm{0.26}$ & $12.18\pm{0.26}$ \\      
    \hline \hline
    \end{tabular}
    \caption{The central values of parameters along with their $1\sigma$ errors from our BAO analyses using $\xi^{\rm hh}$, $\xi^{\rm qq}$ and $\xi^{\rm qh}$ for the case of 10 density bins, after marginalizing over other parameters. Note that $\xi^{\rm qq}$ and $\xi^{\rm qh}$ of the same DS bins are combined into a single data vector for the fitting. The errorbars present the statistical errors of a survey volume of 100~(Gpc/$h$)$^3$. The results in this table are to be compared with the results for the case of 5 density bins shown in Table~\ref{tab:mcmc_results}.
    }
    \label{tab:mcmc_results_m10}
\end{table}

\clearpage

\bibliography{bibfile}{}
\bibliographystyle{aasjournal}

\end{document}